\documentstyle[12pt,psfig]{article}
\def\bi{\bibitem}
\newcommand{\be}{\begin{equation}}
\newcommand{\ee}{\end{equation}}
\newcommand{\beq}{\begin{eqnarray}}
\newcommand{\bec}{\begin{center}}
\newcommand{\eec}{\end{center}}
\newcommand{\eeq}{\end{eqnarray}}
\newcommand{\bear}{\begin{array}}
\newcommand{\ear}{\end{array}}

\begin{document}
\title{The CWKB Method of Particle Production in Periodic Potential}
\author{S. Biswas$^{*a)}$, P. Misra $^{a)}$and I. Chowdhury$^{a)}$\\
a) Department of Physics, University of Kalyani, West Bengal,\\
India, Pin.- 741235\\
$*$email:sbiswas@klyuniv.ernet.in}
\date{}
\maketitle
\begin{abstract}
In this work we study the particle production in time dependent periodic potential using the method of complex time WKB (CWKB) approximation. In the inflationary cosmology at the end of inflationary stage, the potential becomes time dependent as well as periodic. Reheating occurs due to particle production by the oscillating inflaton field. Using CWKB we obtain almost identical results on catastrophic particle production as obtained by others. 
\end{abstract}
\section{\bf{INTRODUCTION}}
The importance of particle production in expanding universe has been a matter of interest and investigation since long times.
 Recently the catastrophic particle production has gained importance espcially in discussing the theory of reheating after inflation 
due to oscillating inflaton field. The literature on the subject, \cite{kls:prl,adl:ppic,adl:pl,ast:prl,dl:pl,gkls:prd,kls:prd,kof:hep-ph,
kof:astro-ph,my:ptp,fkyy:prd} though has contradictory statements, has been extensively reported in ref.[6,7]. The mechanism
 behind the catastrophic particle production is now called parametric resonance and stochastic resonance. A beautiful 
 discussion of classical parametric resonance will be found in Landau and Lifshitz \cite{ll:mec}. The recent attempts try to cast 
the equation of motion of the particle (that are produced) in the form of the Mathieu equation \cite{maclac:tamp} starting with 
a basic model describing the inflaton field $\phi$ with a potential $V(\phi)$ interacting with a scalar field $\chi$ that are produced in 
the description. Earlier attempts to treat parametric resonance to explain reheating after inflation were due to Dolgov and Kirilova \cite
{dk:sjn,tb:prd} but in [15] the treatment was not rightly placed.
\par
From a few years back we are advocating \cite{bis1:pramana,bis2:pramana,bis3:cqg,bis4:cqg,bis5:prd,bis6:grg} a method to treat particle production in curved spacetime. The method is called complex trajectory WKB (CWKB) approximation. The method is also applied to the construction of wavefunction of the universe in quantum cosmology \cite{bis5:prd,bis6:grg} with remarkable success and is very transparent from the standpoint of physical arguments. As mentioned already, due to differences in the growth rate factor even in the recent works [9,10,11], we take up the present work to study the reheating mechanism in inflationary cosmology through the method of CWKB and compare our results with the others.
\par
The basic principles and the mechanism inherent in CWKB are discussed in [13,14,15]. We discuss this in section 2 as a preparatory to the following sections. In section 3 we use the CWKB method to study particle production in the large amplitude region. This section will help us understand the analytic theory of parametric resonance which is discussed in the section 4. We synopsize our findings in the concluding section. 
\par
The basic idea that reheating occurs due to particle production by the oscillating inflaton field was proposed by Linde \cite{adl:pl}. The particles so produced through interaction among themselves attain a state of thermal equilibrium with some temperature $T$. This process continues so long the scalar field transfers its all energy to the already produced elementary particles. The temperature at this stage is termed as reheating temperature, $T_r$. A detailed discussion in this respect will be found in the beautiful papers \cite{kof:hep-ph,kof:astro-ph} and references of previous works will also be found in that reference.
\par

Consider the decay of a scalar field (supposed to be inflaton) $\phi\rightarrow \chi\,\chi$. In the system if already there are many $\chi$ particles with $n_k > 1$, then the probability is greatly enhanced due to Bose statistics. For fermion decay we will find that such enhancement would occur due to multiple reflctions and will be explained in a separate publication. Because of periodicity emerging through the inflaton field $\phi$  the system evolves with explosive particle production. The temporal equation of $\chi$ in a flat Friedman background with scale factor $a(t)$ is given by
\be 
\ddot{\chi}_k+3\frac{\dot{a}}{a}\dot{\chi}_k+\left(\frac{\vec{k}\,^2}{a^2}+m_{\chi}^2(0)-\xi R + g^2\phi^2\right)\chi_k=0.
\ee
Here the time dependence of the background field $\phi$ and the scale factor is obtained from the evolution equations
\beq
H^2=\frac{8\pi}{3M_p^2}(\frac{1}{2}\dot{\phi}^2 +V(\phi)),\\
\ddot{\phi}+3H\dot{\phi}+\frac{\partial V}{\partial \phi}=0,
\eeq
where $H= {\dot{a}\over a}$,  $V(\phi)$ is the effective potential of the scalar field. In a particular model, we can parametrize $\chi$ equation in Minkowski space
($a(t)=1$) as
\be
\ddot{\chi}_k+\left(k^2+g^2\sigma^2+2g^2\sigma\, \Phi\, sin\,mt\right)\chi_k=0,
\ee
where the periodicity $\phi(t)=\Phi\, sin\,mt$ comes from the inflaton field.
This equation is now of the form
\be
\ddot{\chi}_k+\omega_k^2(t)\chi_k=0,
\ee
where the time dependent frequency is $\omega_k^2(t)=k^2+g^2\sigma^2+2g^2\sigma sin\,mt$. This periodicity is in the root for vigorous particle production and  show parametric resonance for modes with certain values of $k$. The  equation. (4) can be cast in Mathieu equation form
\be
\chi^{\prime\prime} _k +(A_k-2qcos2z)\chi_k=0,
\ee
with $mt=2z-\pi/2,\,A_k=4\frac{k^2+g^2\sigma^2}{m^2},\,q=\frac{4g^2\sigma\Phi}{m^2},\,$. Here prime denotes differentiation with respect to $z$. The properties of Mathieu equation shows that within the set of resonance bands of frequencies the modes grow as $\chi_k\propto exp(\mu^{(n)}_kz)$ so that it corresponds to exponential growth of occupation numbers of quantum fluctuations as $n_k(t)\propto exp(2\mu^{(n)}_kz)$ and is interpreted as catastrophic particle production. We will now study the equation (6) to develop an analytic theory of parametric resonance. Before that we review the basic principles of CWKB.
\section{\bf{BASIC PRINCIPLES OF CWKB}}
\begin{figure}[h]
\psfig{figure=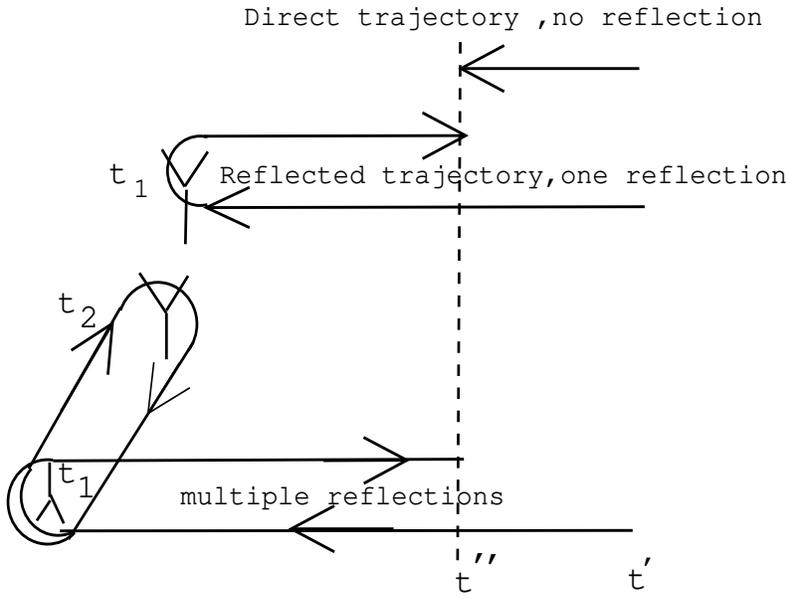,width=13cm,height=12cm}
\caption{Trajectories with no reflection, one reflection and multiple reflections.}
\end{figure}

Consider the eqn.(5) and assume that $\omega(t)$ is nowhere zero for real $t$ but when considered as a function complex $t$ it has complex turning points given by $\omega(t_{1,2})=0$. According to the CWKB, the classical paths contributing in a complex semiclassical approximation joining two prescribed real points $t^{\prime}$ and $t^{\prime\prime}$ (see Fig. 1) are composed of two parts coming from the contribution of the direct trajectory and the reflected trajectory. With 
\be
S(t_f,t_i)=\int_{t_i}^{t_f}\omega(t)\,dt,
\ee
the direct trajectory contribution is written as (we assume that $t^{\prime}>>t^{\prime\prime})$ 
\be
D.T\equiv \frac{1}{\sqrt{\omega(t)}}e^{iS(t^{\prime\prime},t^{\prime})}.
\ee
The reflected trajectory starts from $t^{\prime}$ and moving backward in time gets reflected from the turning point $t_1$ and moving forward arrives at $t^{\prime\prime}$. This contribution is then multiplied by $0,1,2,3,...$ reflections between the the two complex turning points $t_1$ and $t_2$. This contribution is written as 
\be
R.T\equiv-i\frac{1}{\sqrt{\omega(t)}}e^{iS(t_1,t^{\prime})-iS(t^{\prime\prime},t_1)}\sum_{\mu=0}^{\infty}\left[-ie^{iS(t_1,t_2)}\right]^{2\mu}. 
\ee
Using
\be
\sum_{\mu=0}^{\infty}\left[-ie^{iS(t_1,t_2)}\right]^{2\mu}=\frac{1}{1+exp(2iS(t_1,t_2))},
\ee
and taking $t^{\prime\prime}=t$ we get with the replcement $t^{\prime}\rightarrow \infty$ 
\be
\chi_k(t,\infty)\rightarrow e^{iS(t,\infty)}+iRe^{-iS(t,\infty)},
\ee
where we have neglected the WKB pre-exponential factor for convenience. The reflection amplitude is given by
\be
R=-\frac{e^{2iS(t_1,\infty)}}{1+e^{2iS(t_1,t_2)}}.
\ee
Using Feynman-Stuckleberg prescription and the boundary condition that there is no particle at $t\rightarrow -\infty$ i.e.,
\be
\chi_k(t\rightarrow -\infty)\sim Te^{iS(t,t_0)},
\ee
we identify
\be
R_c=-\frac{e^{2iS(t_1,\infty)}}{\sqrt{1+e^{2iS(t_1,t_2)}}}=\sqrt{1+e^{2iS(t_1,t_2)}}R,
\ee
 as the pair production amplitude with $|R_c|^2+|T_c|^2=1$. In (12) and (14) $R$\,($\equiv$ full S-matrx element) is related to 
disconnected propagator and $R_c$ ($\equiv$ connected S-matrix element) is related to connected propagator [19]. The boundary 
conditions (11) and (13) are known as scattering boundary condition and is applied in time dependent gauge. 
\par
Hence for certain complex values of multiple reflection terms, (14) may exhibit poles for some parameter values if
\be
S(t_1,t_2)=(N+\frac{1}{2})\pi
\ee
with $N $ an integer. When the conditions (15) and $\omega(t_{1,2})=0$ are not satisfied or the reality condition on $z$ are violated, the physical-region pole becomes a resonance. This resonance occurs in parameter space and we will call it parametric resonance. The poles add up to give nonperturbative contributions enhancing $R_c$. Thus in CWKB we have a transparent idea of resonance particle production. Depending upon the parameter values in (15) (i.e., the values in $A_k$ and $q$ in eqn.(6)) we have resonance particle production in CWKB. In the next section we will discuss the particle production in large amplitude region.
\section{\bf{PARTICLE PRODUCTION IN THE LARGE\\ AMPLITUDE REGION}}
In the large amplitude region the parameter $q$ is large (see eqn.(6)) and hence the resonance is broad. In that case
 we replace the periodic barrier by piece-wise inverted harmonic oscillators and  within the 
interval $|z| << \pi/4$ we write $cos(2z)\simeq 1-2z^2$. Let us try to understand this 
replacement. Considering (6) as a Schr\"{o}dinger-like equation
not in space but in time, in the potential problem $qcos2z$ behaves as potential energy 
and $A_k/2$ as the total energy. For $A_k<<\mid q\mid $, there is an infinite succession of 
large potential barriers separated by allowed region given by $A_k-2qcos\,2z>0$. Instead of considering the barriers 
at once, we consider a single barrier and 
calculate 
the particle production which occurs at the turning points given by 
\be
\omega(z)\equiv A_k-2qcos2z=0
\ee
 These turning points occur at 
\be
z\simeq (\frac{(2n+1)\pi}{4}+(-1)^n\frac{A_k}{4q}), \rm with\,\, n=0,\pm 1, \pm 2, \pm3, ... 
\ee
for large $q$. Thereafter we repeat the process $n$ times with $n\simeq z/ \pi$, $n$ being a large integer. Now to calculate the 
particle production we are to determine particle and antiparticle states i.e., the left and right moving mode solutions with respect to
 which the {\it in} vacuum is defined. 
\par
In CWKB these should be the adiabatic modes at large $z$. In other words  the mode solution should satisfy the adiabaticity 
condition. As we have repeated barriers we are to find out the adiabatic modes at finite $z=z_0$ such that the modes effectively 
describe the vacuum  at finite $z$, for large $q$.  Let us consider the central barrier with $-\frac{1}{4}(\frac{\pi}{4}+\frac{A_k}{q})
<z<\frac{1}{4}(\frac{\pi}{4}-\frac{A_k}{q}) $. For large $q$ the approximation gets better and the adiabatic condition is satisfied at finite 
$z=z_0$ for $\mid z\mid \ll \pi/4$  but far away from the turning point $z=\frac{1}{4}(\pi-\frac{A_k}{q})$ (considering a single barrier). 
In such a case the above replacement is a good approximation. As we are considering adiabatic modes far away from the turning
 points (basically these are the WKB modes),  the reflection and transmission coefficients or the Bogolubov coefficients will be 
constant  for a barrier. The maximum of potential barriers occurs at $\pi, 2\pi, 3\pi, ...$ and hence the resonance occurs for a broad
 range of values of $k$ with $k^2/m^2=A_k-2q$. Now the resonance width varies as $q^l$ with $l=1, 2, 3, ... $ . So for  narrow
 resonance, the first band with  $l=1$ will be important with $A_K=1\pm q.$  Let us consider the broad resonance. In this section 
we consider only the production of particles at each barrier as if there is no particles in the previous barrier and also neglect the 
enhancement at the zeroes of the inflaton field where non adiabatic transitions occur . 
Introducing $cos2z\simeq 1-2z^2$ and with
\be
z^{\prime}=(4q)^{1/4}z,\,\,\lambda=\sqrt{q}\epsilon,\,\,\epsilon=\frac{A_k}{2q}-1,
\ee
the Mathieu equation (6) reduces to the form
\be
\frac{d^2\psi}{{dz^{\prime}}^2}+(\lambda +{z^{\prime}}^2)\psi=0.
\ee
The left and right moving modes having the asymptotic behaviour of unit flux [10] are given as,
\bec
$\chi_{L,R}=(2\sqrt{q}z)^{(-1\mp i\sqrt{q}\epsilon/2)}exp(\mp i\sqrt{q}z^2)$ 
\eec
which can be obtained from the solutions of (19) written in terms of parabolic cylinder functions as
\bec
 $\chi_L(z)\propto  D_{-(i \sqrt{q}\epsilon +1)/2}(e^{i\pi/4}2q^{1/4}z)$
 \eec
with
\bec
 $\chi_R{(z)}=\chi_L^*(z)$.
\eec
 Despite the finite range $z_0<<\pi/4$, the large $q$ in the argument of D-function makes it possible to use the asymptotic form 
of these D-functions to use as mode solution as if we are considering the WKB adiabatic modes.
 In CWKB we do not require the exact mode solutions to evaluate the particle production amplitude; however we are to guarantee 
that the in vacuum is rightly specified. To evaluate $R$ as in (12) we are to evaluate the complex turning points.
The complex turning points of (19) are at $z^{\prime}=\pm i\lambda$. Using CWKB expression of reflection amplitude, we find after the 
evaluation of the integrals
\beq
|R_c|^2 & = & \frac{e^{-\pi\lambda}}{1+e^{-\pi\lambda}}\,,\nonumber\\
      & = & \frac{e^{-\pi\sqrt{q}\epsilon}}{1+e^{-\pi\sqrt{q}\epsilon}}\,.
\eeq    
When we consider full Mathieu equation, the turning points will be repeated and the solution will be of the form
\be
\psi(z)\sim |R|^n P(z)\equiv e^{\lambda z}\,P(z)=e^{nln|R|}P(z),
\ee
where $n\simeq z / \pi$ is the number of barriers crossed during the time $z$. Hence 
\beq
\lambda & = &  \frac{1}{\pi}\,ln|R_c|,\nonumber\\
        & = &  \frac{1}{\pi}\,ln\,\frac{exp(-\pi\sqrt{q}\epsilon/2)}{1+exp(-\pi\sqrt{q}\epsilon)}.
\eeq
If $q$ is large, we get
\be
\lambda \simeq -\sqrt{q}\epsilon/2
\ee

This result exactly coincides with the result obtained by Fujisaki et. al.\cite{fkyy:prd}. This result amply reflects the usefulness 
of the complex turning points in the CWKB method. The construction of CWKB wavefunction as $\sim |R|^n$ is however a crude 
estimate. In obtaining (23), the multiple reflections between the complex turning points have been ignored and the periodic aspect
 of the potential has been introduced in a qualitative way through (21). However, a general treatment, taking the effect of repeated 
barriers, will now be taken within the framework of CWKB. 
\par
It should be pointed out that while calculating the reflection potential 
as $\mid R\mid^n$ we basically assume that (i) there is no particles produced in the previous barrier , (ii) and neglect the non 
adiabatic changes at the zeroes of the inflaton field and the phases that the field acquire before entering a potential barrier due to scattering. 
\par 
It should be pointed out that in the above discussion we have obtained the reflection and transmission coefficients as constants 
when we consider adiabatic evolution far away from the turning points that lie around the vicinity of the zeroes of the inflaton field. 
To take into account the non adiabatic changes that occur around the zeroes of the inflaton field,  we will  expand  $\omega^2(t)$ around the zeroes of the inflaton field.
For the purpose we would bring $\omega^2(t)$  term in the form $A+ B sin^2mt$ and then expand around the zeroes $t=t_j$ of the inflaton field. In 
this case also we can safely assume the adiabatic evolution in the region away from the zeroes $t=t_j$ of the inflaton field. We can then 
take the coefficients 
of the CWKB adiabatic modes $exp(\pm i\omega(t))$ as constants until we face the next barrier where the coefficients will change.
 We take up all these aspects in the next section. 

\section{\bf{ANALYTIC THEOTY OF PARAMETRIC\\ RESONANCE}}
\begin{figure}[h]
\psfig{figure=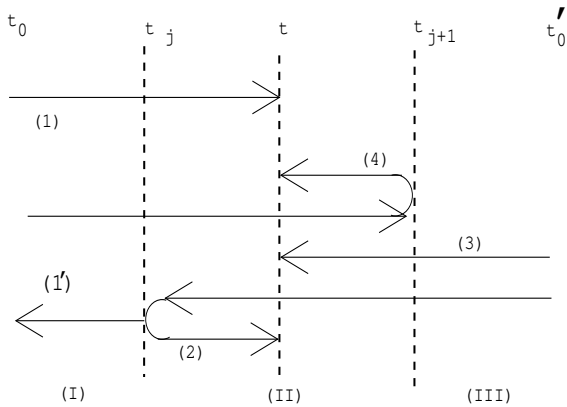,width=10cm,height=8cm}
\caption{The CWKB trajectories in periodic potential}
\end{figure}
Our objective in this section is to develop  an analytic theory of parametric resonance particle production using the method of CWKB.
We consider a Schrodinger equation with time dependent potential
\be
\frac{d^2\psi}{dt^2}+[k^2-V(t)]\psi= 0,
\ee
where the periodicity in the potential is given by $V(t+T) = V(t)$, where $T$ is the period. In the previous section we have not 
considered the effect of periodicity explicitly. When $V(t)$ is periodic the situation is quite interesting. We now have two types 
of reflection points, one is complex and other is real in time. For the latter we will use tunneling boundary condition akin to 
particle production in space dependent gauge \cite{bis7:grg,bis8:ijmp}. Suppose $V(t)=V_0\, cos2t$ with period $T=\pi$. The 
solution of (24) will then behave as $exp\,[\pm(k^2-V_0)^{1/2}]$ as if at these points the solutions undergo non-adiabatic changes. 
This implies that between two such reflection points, as clarified in the previous section, the solution undergoes adiabatic evolution 
where the WKB approximation can safely be used. We now call the points $t_j$ where $V(t_j)=V_0$ as `reflection points' and the 
complex points where $k^2-V(t)=0$ as the `turning points'. The inflaton $\phi(t)$ which generates this $V(t)$ is zero at these points. (see (4)).
 There is another aspect of periodicity. In a particular barrier, between $t_j$ and $t_{j+1}$, there will be some particles already 
produced with respect to the previous barrier and hence the boundary condition of `no particle state' at $t\rightarrow -\infty$ will not
 be satisfied. This fact is to be taken into account while considering particle production in a given barrier.   
\par
Let us define
\be
\omega(t)\,=\,[k^2-V(t)]^{1/2}
\ee
and
\be
S(t_f,t_i) = \int_{t_i}^{t_f}\,\omega(t)\,dt
\ee
as the action for the path that goes from $t_0$ to $t$. Suppose we have two complex turning points $t_1$ and $t_2$ given by 
$\omega(t_{1,2})\,=\,0$. We adopt the boundary conditions such that at $t\rightarrow -\infty$ we have the transmitted wave
\be
\psi(t\rightarrow -\infty)=T_ce^{iS(t,t_0)},
\ee
and at $t\rightarrow +\infty$
\be
\psi(t\rightarrow +\infty)=e^{iS(t,t_0)}+R_ce^{-iS(t,t_0)}.
\ee
Here $R_c$ and $T_c$ are respectively the reflection and the transmission amplitude. In CWKB the expression of the reflection amplitude is given in (14)
and $T_c$ is determined from the relation $|R_c|^2+|T_c|^2 = 1$. Equation. (27) and (28) have the interpretation that at $t\rightarrow -\infty$ we have no particle but at $t\rightarrow +\infty$ we have pair production in the {\it{out}} vacuum with respect to {\it{in}} vacuum. The same problem can be evaluated in terms of Bogolubov mode decomposition technique with
\be
\psi(t\rightarrow +\infty)=\alpha_{\omega}e^{iS(t,t_0)}+\beta_{\omega}e^{-iS(t,t_0)},
\ee
in which the the Bogolubov coefficients are given in terms of transmission and reflection coefficients as [20]
\be
|\alpha_\omega|^2 = \frac{1}{|T_c|^2},\,\,|\beta_\omega|^2 = \frac{|R_c|^2}{|T_c|^2}
\ee
In the above treatment the particle and antiparticle states are defined at $t\rightarrow \pm\infty$ with the corresponding vacua respectively as {\it{out}} and {\it{in}} vacuum.   
We now proceed toward the construction of CWKB eigenfunction. Let $t_j$ and $t_{j+1}$ be the two points where $V(t_j)=0$ and $t$ is a point such that $t_j<t<t_{j+1}$ where we want to calculate the eigenfunction. We choose the rightmoving and left moving waves as follows (see fig.2).
\begin{center}
Rightmoving\,\,wave\,\,from\,\, $t_0$\,\,to\,\,$t=exp\,[-iS(t,t_0)]$\,,\\
Leftmoving\,\,wave\,\,from\,\,${t_0}^\prime$\,\,to\,\,$t =exp\,[iS(t,{t_0}^\prime)]\,,$
\end{center}
where $S(t_i,t_f)$ is given in eqn.(26). For the rightmoving wave the transmission and the reflection coefficients are $(T_k,R_k)$ 
whereas for the left moving wave we take the coefficients as $(T_k^*,R_k^*)$. We now avoid the subscript $c$ on $R$ and $T$. Let
 in the region I, $t_{j-1}<t<t_j$, the CWKB solution before tunneling is (represented by (1) and ($1^\prime$) in Fig. 2)
\be
\psi_k^j(t)=\frac{\alpha_k^j}{\sqrt{2\omega}}e^{-iS(t,t_0)}+\frac{\beta_k^j}{\sqrt{2\omega}}e^{+iS(t,t_0)},
\ee
The Bogolubov coefficients in a given barrier are now supposed to be constants.  
Between $t_j<t<t_{j+1}$, in the region II we take similarly 
\be
\psi_k^{j+1}(t)=\frac{\alpha_k^{j+1}}{\sqrt{2\omega}}e^{-iS(t,t'_0)}+\frac{\beta_k^{j+1}}{\sqrt{2\omega}}e^{+iS(t,t'_0)},
\ee
where the coefficients $\alpha_k^{j+1}$ and $\beta_k^{j+1}$ are constants for $t_j<t<t_{j+1}$. Henceforth we will not write 
the WKB pre-exponential factor for convenience. The constuction of $\psi_k^{j+1}(t)$, according to CWKB is now shown in the 
fig. 2. In the region $t_j<t<t_{j+1}$ the rightmoving part consists of two parts. The part (1) represented by $\alpha_k^j\, exp\,[-iS(t,t_0)]$
 after transmission gets multiplied by $1/{T_k} $ because of transmission at $t_j$ and using (30) we find
\be
\alpha_k^j \,exp\,[-iS(t.t_0)]\rightarrow \frac{\alpha_k^j}{T_k}\,exp\,[-iS(t,t_0)].
\ee     
There is also a contribution to the rightmoving wave coming from the region III where $t>t_j$, as shown in the fig.2, giving the part $\beta_k^j exp[+iS(t,t_0)]$ in the region $t<t_j$ represented by $(1')$ in the figure 2. The amplitude of the leftmoving part at $t_j$ is $\beta_k^j\, exp\,[+iS(t_j,t_0)]$. This amplitude part when continued in the region $t>t_j$ becomes $\beta_k^j/T_k^*$ which again being reflected at $t_j$  becomes $\beta_k^jR_k^*/T_k^*$ so that in the region $t>t_j$ 
\be
\beta_k^j\,exp\,[iS(t,t_0)]\rightarrow \frac{\beta_k^j\,R_k^*}{T_k^*}\,exp\,[+iS(t_j,t_0)]\,exp\,(-iS(t,t_j)).
\ee
Writing $S(t,t_j)=S(t,t_j)+S(t_j,t_0)-S(t_j,t_0)=S(t,t_0)-S(t_j,t_0)$, we get from (34) for the rightmoving part as
\be
\psi_{k,RMP}^j=\left[\,\frac{\alpha_k^j}{T_k}+\frac{\beta_k^j\,R_k^*}{T_k^*}\,e^{2\,iS(t_j,t_0)}\,\right]e^{-iS(t,t_0)}.
\ee
To get (35) we can also start from the region III with $\beta_k^{j+2}exp(iS(t,t'_0))$ that now gets multiplied by $R_k^*/T_k^*$ 
because of transmission at $t_{j+1}$ and reflection at $t_j$ and then use the continuity condition 
\begin{center}
$\beta_k^j e^{iS(t_j,t_0)}=\beta_k^{j+2}e^{iS(t_j,t'_0)}.$
\end{center}
We will get the same result as (35). Comparing (35) with the first term in (32) we get
\be
\alpha_k^{j+1}=\left[\,\frac{\alpha_k^j}{T_k}+\frac{\beta_k^j\,R_k^*}{T_k^*}e^{2\,iS(t_j,t_0)}\,\right].
\ee
Now the left moving part has two contributions given by the trajectory (3) and (4) of fig. 2. The trajectory (3) coming from $t_0^\prime$ gives the left moving part $\beta_k^{j+2}\,exp\,(iS(t,t'_0))$ in the region $t<t_{j+1}$. Now we use the relation above to convert $\beta_k^{j+2}$ in terms of $\beta_k^j$  This part then in $t>t_j$ becomes
\be
\frac{\beta_k^j}{T_k^*}\,e^{iS(t,t_0)}.
\ee
Another contribution comes from $\alpha_k\,exp\,(-iS(t,t_0))$ which on transmission at $t_j$ and reflection at $t_{j+1}$ gets modified to
\be
\frac{\alpha_k^j}{T_k}R_ke^{-iS(t_{j+1},t_0)+iS(t,t_{j+1})}
\ee
Now, $iS(t,t_{j+1})=iS(t,t_{j+1})+iS(t_{j+1},t_0)-iS(t_{j+1},t_0)=iS(t,t_0)-iS(t_{j+1},t_0)$ so that the left moving part becomes
\be
\psi_{k,LMP}^{j+1}=\left[\,\frac{\beta_k^j}{T_k^*}+\frac{\alpha_k^j\,R_k}{T_k}e^{-2iS(t_{j+1},t_0)}\,\right]e^{+iS(t,t_0)}.
\ee
Hence comparing (39) with the second term in (32) we get
\be
\beta_k^{j+1}=\left[\,\frac{\beta_k^j}{T_k^*}+\frac{\alpha_k^jR_k}{T_k}e^{-2iS(t_{j+1},t_0)}\,\right].
\ee
This result exactly coincides with Kofman, Linde and Starobinsky \cite{kls:prd}. We have in the expression of $\beta_k^{j+1}$ the phase term as $S(t_{j+1},t_0)$ instead of $S(t_{j},t_0)$. It should be pointed out that we have not taken repeated reflections between the turning points $t_j$ and $t_{j+1}$ which will automatically be introduced when we calculate $R_k$ and $T_k $ through the technique of CWKB.
\par

Now we are to determine the transmission and reflection coefficients using CWKB in the region $t_j<t<t_{j+1}$. For the purpose we need to specify $V(t)$. We take $V(t)= -g^2\Phi^2\,sin^2(mt)$. Actually this type of terms arise in inflationary scenario from an effective potential $V(\phi)=\frac{m^2}{2}\phi^2$, and the interaction term $-\frac{1}{2}g^2\phi^2\chi^2$. Instead of (4) we now have
\be
\ddot{\chi}_k\, +\,(k^2\,+\,g^2\,\Phi^2\,sin^2(mt))\chi_k=0
\ee
In the vicinity of $t_j$ equation (41) is transformed to
\be
\frac{d^2X_k}{dt^2}\,+\,\left(k^2\,+\,g^2\Phi^2m^2(t-t_j)^2\right)\,X_k=0.
\ee
With $\tau=k_*(t-t_j),\,\lambda=\frac{k}{k_*}\,,\,k_*^2=g\Phi m$, (38) reduces to
\be
\frac{d^2X_k}{d\tau^2}\,+\,(\lambda^2+\tau^2).
\ee
Using the results in (18) and (28) we find
\beq
R_k & = & \frac{-ie^{-i\phi_k}}{\sqrt{1+e^{\pi\lambda^2}}}\,,\\
T_k & = & \frac{-ie^{-i\phi_k}}{\sqrt{1+e^{-\pi\lambda^2}}}.
\eeq
In CWKB the phase $\phi_k$ is unknown. However for the particular problem it can be calculated knowing the solutions of (39) in terms of parabolic cylinder functions and we get
\be
\phi_k= arg\Gamma\left(\frac{1+i\lambda^2}{2}\right).
\ee
We now calculate the number density of outgoing particles as $n_k^{j+1}=|\beta_k^{j+1}|^2$ and find after simple algebra
\beq
n_k^{j+1} & = & e^{-\pi\lambda^2}\,+\,(1+2e^{-\pi\lambda^2})n_k^j\nonumber\\
          &   & -2e^{-\frac{\pi}{2}\lambda^2}\sqrt{1+e^{-\pi\lambda^2}}\sqrt{n_k^j(1+n_k^j)}\,sin\theta^j_{tot},
\eeq
where the phase $\theta^j_{tot}=2\theta^{j+1}_k-\phi_k+arg\beta_k^j-arg\alpha^j_k$. In obtaining (47) we have used $|\alpha_k^j|-|\beta_k^j|^2=1$ and$ |\alpha_k^j|=\sqrt{1+n_k^j}).$ Here the arbitrary point $t_0$ is now taken as zero and
\be
\theta_k^{j+1}=S(t_{j+1},0)=\int_0^{t_{j+1}}[(k^2+g^2\Phi^2sin^2(mt))]^{1/2}dt.
\ee
For narrow resonance i.e.,$q=\frac{g^2\Phi^2}{4m^2}<<1$ and with $A_k=\frac{k^2}{m^2}+2q$ we will find
\be
S(t_{j+1},t_j)\simeq\sqrt{A_k}\pi=n\pi,
\ee
taking into consideration the effect of periodicity i.e., at the points $t_j$ the phase in (51) must be an integral multiple of $\pi$. We 
have also neglected the terms like ${\it{O(q/A_k)}}$ while evaluating the right hand side of (48) through elliptic integrals. Thus
\be
e^{2i\theta_k^{j+1}}=(-1)^ne^{\theta_k^j},
\ee
and our result then exactly coincides with the results given in [6,7,8]. Apart from this minor difference
all the results arrived in [6,7,8] will also be the same in our case. Our CWKB treatment in this section and in the previous 
section (eqns. (22) and (23)) simply reflects the region where both the results are valid.  
\par
Let us briefly mention the important results obtained from analytical analysis.\\
\medskip
(i) In CWKB the resonance particle production occurs due to rotation of currents at the `turning points' and `reflection points'. The former is independent of time and we call it spontaneous particle creation and the latter is due to multiple reflections at the points $t_j$ and we call it induced particle creation.\\
\medskip
(ii) Due to spontaneous particle creation the number of produced particles always increases whereas the induced creation may result in a destructive interference between $t_j$ and $t_{j+1}$ when $sin \theta _j$ remains positive and varies. This solely occurs due to the time dependence contained in $\theta_j$ and multiple reflections between $t_j$ and $t_{j+1}$.\\

\medskip
(iii) Actually the spontaneous particle creation  determines the resonance structure. A large $\pi\lambda^2 $ will suppress the the effect of particle creation. Hence we must have
\be
\lambda^2=\frac{A_k-2q}{2\sqrt{q}}\leq \pi^{-1}
\ee
where $A_k= \frac{k^2}{m^2}+2q\,,\,q=\frac{g^2\Phi^2}{4m^2}.$ This gives an estimate of resonance width as
\be
k^2\leq k_*^2=gm\Phi/ \pi.
\ee
(iv) Now $exp(-\pi\lambda^2)\propto exp(-1/g)$ and is non-analytic at $g=0.$ Thus (45) manifests the non-perturbative nature of CWKB resonance effects.\\
\medskip
(v) In the broad resonance regime $q >> 1$. We take $\Phi(t)=$ constant; it makes $\theta_k$ and $\phi_k$ independent. It can be shown that in that case the analytical solution nicely matches with the numerical solution of Mathieu equation. The details will be found in [6,7,8].
\par 
Let us consider an interesting interpretation of  CWKB reflection coefficient. From the expression (12), $|R|^2$ has the following 
heuristic interpretation. Since
\be
|R|^2 =\frac{1}{n+1} \frac{n}{(n+1)},
\ee 
and $1/{(n+1)}$ is the probability that {\it{in}} vacuum evolves into the {\it{out}} vacuum, we say that out of $n+1$ particles in 
the $|out>$ vacuum, $n$ particles get reflected so that $|R|^2$ gives the absolute probability of one pair production from 
the {\it{in}} vacuum as if $n $ paricles remain in the $|in>$ the vacuum. Hence $(1+n)^2|R|^2$ gives the number of produced 
pairs in the $|out>$ vacuum. Now in CWKB $exp(-\pi\lambda^2)$ is the number of particles produced in any barrier. The number 
 $n_k^j$ particles already are in the in vacuum, so we do not require the first factor in (53). Thus
\be
|R|^2=\frac{n_k^j}{n_k^j+1}n_k^j\simeq n_k^j.
\ee

Hence
\be
(1+e^{-\pi\lambda^2})^2\,n_k^j
\ee
gives as the number of particles produced in the $out$ vacuum while $n_k^j $ remaining in the $in$ vacuum. We recall that
 \cite{bis8:ijmp}   
\be
|R|^2=e^{-ln(1+N(k))}|R_c|^2=e^{-2Im {\cal{L}}_{eff}(k)V_4}|R_c|^2.
\ee
When no particles are produced $N(k)=0$, we get $|R|=|R_c|$ and $Im{\cal{L}}_{eff}=0$ leading to $|in> = |out>.$  In that case 
we have two possibilities: (i) the initial vacuum contains no particles and hence $|R|=|R_c|=0$, (ii) the initial vacuum contains 
some particles (squeezed $|in>$ state) then from (12), 
$(1+exp(2iS(t_1,t_2)))^2|R|^2=(1+exp(2iS(t_1,t_2)))^2|R_c|^2=(1+exp(2iS(t_1,t_2)))^2|n|^2$
gives the number of particles in the out vacuum due to presence of n particles in the squeezed $|in>$ vacuum. Hence the number 
of particles $n_k^{j+1}$ in the barrier between $t_j<t<t_{j+1}$ due to $n_k^j$ particles in the $|in>$ vacuum is \\
\begin{center}
$n_k^{j+1}=$ particles produced in $out$ vacuum for zero particle $in$ state $+\, (1+exp(2iS(t_1,t_2)))^2|n_k^j|^2).$  
\end{center}     
Hence,
\be
n_k^{j+1}= e^{-\pi\lambda^2}+(1+e^{-\pi\lambda})^2n_k^j
\ee
This result coincides with the first two terms of (47) when $exp(-\pi\lambda^2)<<1$. If $n_k^j=0$ i.e., zero particle in 
the {\it{in}} vacuum we get the standard result $n_k= exp(-\pi)\lambda^2$ for any barrier. The skeptical readers may avoid this 
heuristic arguments.
\par
Next, consider the large occupation number situation in which $n_k>>1$. In that case we write (45) as
\be
n_k^j=n_k^j exp(2\pi\mu_k^j),
\ee
where
\be
\mu_k^j=\frac{1}{2\pi}ln\,(1+2e^{-\pi\lambda^2}-2sin\theta^j_{tot}e^{-\pi\lambda^2/2}\sqrt{1+e^{-\pi\lambda^2}}).
\ee
After a number of inflaton oscillations we write the number of $\chi$ particles occupation number $n_{k}$ as
\be
n_k(t)= \frac{1}{2}e^{2\pi\sum_j\mu_k^j}\simeq e^{2m\int^t\,dt\mu_k(t)},
\ee
so that
\be
n_{\chi}=\frac{1}{4\pi^2}\,\int\,dk\,k^2e^{2m\mu_kt}.
\ee
We now evaluate the integral by steepest descent method with $\mu_{max}=\mu$ at  $k=k_m$ and using $\mu_k^{\prime\prime}\sim 2\mu/{\delta k^2}$, we finally get
\be
n_{\chi}\simeq \frac{1}{8\pi^2}\frac{\delta k\,\,k_m^2\,e^{2\mu mt}}{\sqrt{\pi\mu\,mt}}.
\ee
The above result comes out the same as that obtained in [7].
\section{\bf{CONCLUSION}}
In this work we find that the complex trajectory WKB approximation nicely reproduces all the essential aspects of particle 
production due to inflaton oscillation. The reason of difference with other works [10, 11] is due to not taking the multiple 
reflections (a non-perturbative quantum corrections) in their approach. Except for a phase factor the coincidence of our results 
with that of [6,7,8] leads credence to the CWKB approach. In this work we have concentrated the treatment to the Minskowski 
spacetime only to understand the effect of periodic potential on particle production. With $a(t)\neq 1$, the qualitative conclusion 
will not change remarkably.  In curved spacetime $\omega^2(t)$ will be modified as 
\bec
$\omega^2(t) =\frac {k^2}{a^2(t)}+g^2\Phi^2 sin^2mt +\Delta\omega $
\eec
where
\bec
$ \Delta\omega=m_\chi^2-\frac{3}{4}(\dot{a}/a)^2-\frac{3}{2}(\ddot{a}/a)-\xi R$ 
\eec
is usually very small in the region of inflaton oscillation where $H^2<< m^2$ and can be neglected. In this case we are to replace 
$\chi$ by $X_k(t)=a^{3/2}\chi_k(t)$. In the broad resonance regime the effect of $a(t)$ might result in the number of produced 
particles at some time depending upon the parameter values and the FRW scale factor. The details in this respect will be reported 
elsewhere. Actually in our approach we do not require much to depend on Mathieu equation, everything can be 
settled out from the CWKB framework. The stochastic resonance (in which the number of particles may decrease at some moments),
 backscattering and fermion production can also be taken in our frame work. In our previous works [15, 16] we have discussed 
fermion production without repeated reflections between the `reflection points'. We like to take all these aspects in a future 
publication.   

{\bf{Acknowledgement}}\\
The authors are grateful to Dr. Bijan Modak and Prof. Padmanava Dasgupta for very helpful discussion and useful comments during the preparation of the paper.      
\end{document}